\documentclass[aps,prx,onecolumn,citeautoscript,nofootinbib,preprint]{revtex4}
\usepackage{amsmath}
\usepackage{amssymb}
\usepackage{float}
\usepackage[section]{placeins}
\usepackage{bbm}
\usepackage{bm}
\usepackage{comment}
\usepackage{graphicx}
\usepackage{physics}
\usepackage{color}
\usepackage[papersize={8.5in,11in}]{geometry}
\usepackage[colorlinks=true]{hyperref}
\hypersetup{
    bookmarks=true,         
    unicode=false,          
    pdftoolbar=true,        
    pdfmenubar=true,        
    pdffitwindow=false,     
    pdfstartview={FitH},    
    pdftitle={S},    
    pdfauthor={S. Sachdev},     
    pdfsubject={},   
    pdfcreator={},   
    pdfproducer={}, 
    pdfkeywords={} {} {}, 
    pdfnewwindow=true,      
    colorlinks=true,       
    linkcolor=magenta, 
    citecolor=blue,        
    filecolor=magenta,      
    urlcolor=blue           
}

\usepackage{amsfonts}
\usepackage{upgreek}
\usepackage{slashed}
\usepackage{latexsym}
\usepackage[export]{adjustbox}
\usepackage{dsfont}
\usepackage{subcaption}
\begin{document}

\preprint{\href{https://arxiv.org/abs/2304.13744}{arXiv:2304.13744}}
\title{Quantum statistical mechanics of\\ the Sachdev-Ye-Kitaev model\\ and charged black holes}
\author{Subir Sachdev}
\affiliation{Department of Physics, Harvard University, Cambridge MA-02138, USA}
\date{\today}
\begin{abstract}
This review is a contribution to a book dedicated to the memory of Michael E. Fisher.
The first example of a quantum many body system not expected to have any quasiparticle excitations was the Wilson-Fisher conformal field theory. The absence of quasiparticles can be established in the compressible, metallic state of the Sachdev-Ye-Kitaev model of fermions with random interactions. The solvability of the latter model has enabled numerous computations of the non-quasiparticle dynamics of chaotic many-body states, such as those expected to describe quantum black holes. We review thermodynamic properties of the SYK model, and describe how they have led to an understanding of the universal structure of the low energy density of states of charged black holes without low energy supersymmetry.\\~\\
{\sffamily 
\begin{center} 
Published in\\ \href{https://doi.org/10.1142/13571}{\bf 50 years of the renormalization group},\\ dedicated to the memory of Michael E. Fisher,\\ edited by Amnon Aharony, 
Ora Entin-Wohlman, David Huse, and Leo Radzihovsky,\\ World Scientific.
\end{center}
}
\end{abstract}

\pacs{Valid PACS appear here}
\maketitle{}


\section{Introduction}
\label{sec:intro}

\begin{figure}[t]
\begin{center}
\includegraphics[width=5in]{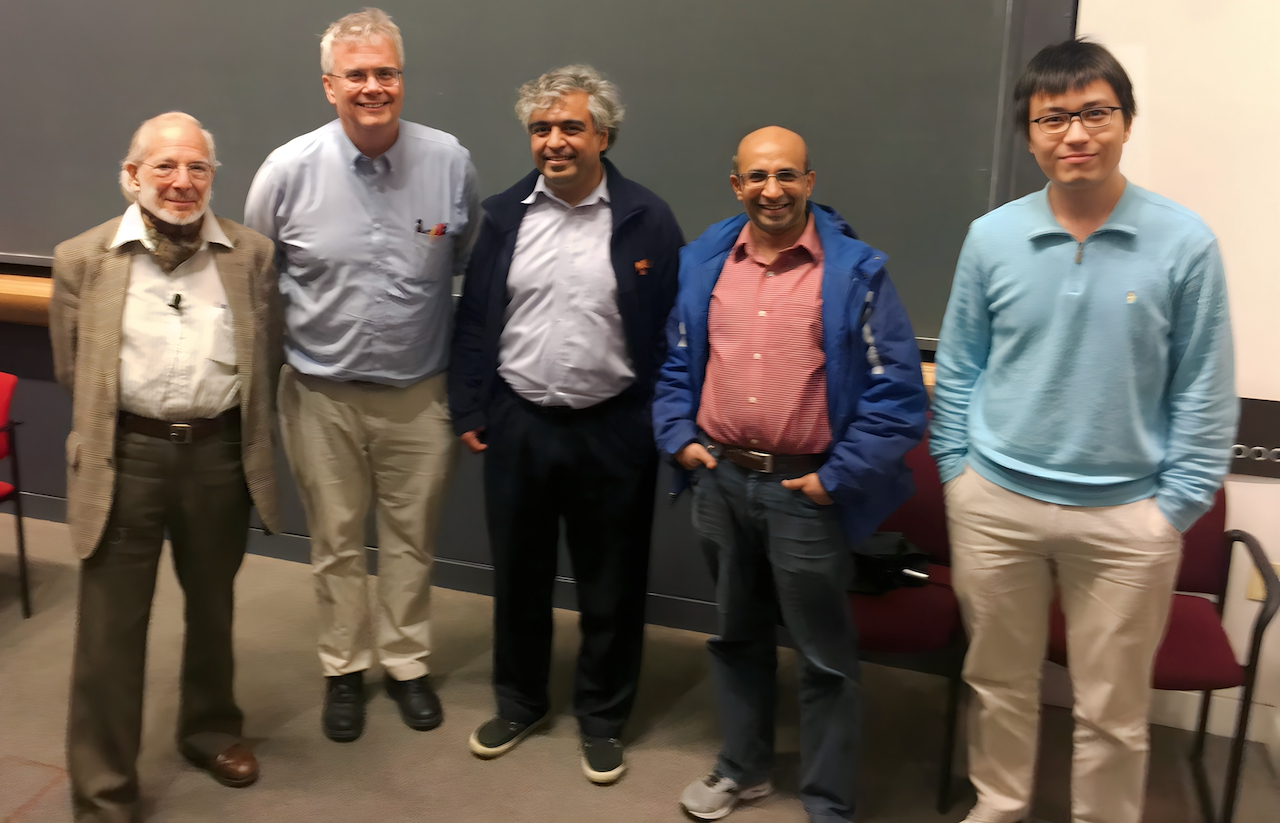}
\end{center}
\caption{Michael Fisher on the occasion of the Lee Historical Lecture at Harvard in 2017. Shown here with 5 generations in the Fisher academic tree: Michael Fisher, David R. Nelson, the author, T. Senthil, and Chong Wang.}
\label{fig:mef}
\end{figure}

The early work of Michael Fisher on critical points in fluid and magnetic systems \cite{Fisher_Colorado} pioneered a line of thought which ultimately transformed the fields of condensed matter physics and high energy particle physics. I recall the strong impression his articles made on me as a young student in the early 80's. The idea of `universality' seemed almost magical: why should an Ising model of magnetic spins on a lattice have the same critical exponents as the liquid-gas critical point of carbon dioxide? The combination of mathematical elegance and experimental relevance was unbeatable, and convinced me to become a condensed matter theorist. I was fortunate to learn more about these topics from my Ph.D. advisor David Nelson, one of Michael Fisher's Ph.D. advisees, and so become part of Michael Fisher's remarkable academic tree (see Fig.~\ref{fig:mef}).

The ideas of universality and universal scaling functions led to the development of the renormalization group, and of renormalization group fixed points, by Fisher, Kadanoff, Wilson, and others. These developments greatly broadened the impact of Michael Fisher's ideas. The Wilson-Fisher fixed point \cite{WilsonFisher} provided the first example of a conformal field theory which was strongly interacting. Earlier examples, such as the Bethe ansatz for the antiferromagnetic Heisenberg spin chain \cite{Bethe} and Onsager's solution of the two-dimensional Ising model on the square lattice \cite{Onsager}, provided fixed points which ultimately permitted a free field description, although the relationship between the free fields and the microscopic degrees of freedom was quite non-trivial. 

In a more recent language, the Wilson-Fisher fixed point provides the first example of a many-body quantum system which is not expected to have any quasiparticle excitations {\it i.e.\/} the two-spatial-dimensional Ising model in a critical transverse field is a quantum system which has no particle-like description of its excitations, even asymptotically. One indication of the absence of quasiparticles is that the retarded dynamic susceptibility of the critical Ising model at momentum $k$ and frequency $\omega$ obeys \cite{SY}
\begin{align}
\chi(k, \omega) \sim \frac{1}{[c^2 k^2 - (\omega + i 0^+)^2]^{1-\eta/2}}\,,
\end{align}
with $\eta$ the anomalous dimension exponent introduced by Fisher and Burford \cite{MEFeta}. A non-zero $\eta$ implies that the spectral density of excitations has a branch cut for $|\omega| > c k$, and there is no pole associated with a quasiparticle. 
This point of view is developed in my first book \cite{QPTbook}, and should be contrasted from that invariably taken in quantum field theory monographs, where the asymptotic scattering states are reduced to free particles via a LSZ reduction. I should note that the non-quasiparticle property of the Wilson-Fisher fixed point has not been proven, although it is highly likely: it is not ruled out that some exotic and non-local quasiparticle basis may be discovered in the future, leading to poles in the associated non-local operator.

Michael Fisher expressed the strongly interacting property by the concept of `hyperscaling' \cite{FisherdeG}: for a quantum critical system in $d$ spatial dimensions with dynamic critical exponent $z$ hyperscaling implies that the entropy density $\mathcal{S}$ has the following dependence on temperature
\begin{align}
\mathcal{S}(T \rightarrow 0) = a_1 T^{d/z} \label{e1}
\end{align}
for some constant $a_1$. For a system with relativistic invariance with a velocity $c$ at low energies, we have $z=1$, and we obtain a length scale 
$\hbar c/(k_B T)$, and then the dimensional constant $a_2$ in 
\begin{align}
\mathcal{S}(T \rightarrow 0) = a_2 k_B (k_B T /(\hbar c))^d \label{e2}
\end{align}
becomes a universal number \cite{FisherdeG,CSY94,Diatlyk:2023msc}, dependent only upon the universality class of the critical theory. The number $a_2$ is closely connected to the `central charge' of conformal field theories. 

Another consequence of hyperscaling, and the associated absence of quasiparticles, is that the dynamics of such systems is `Planckian', an idea finding much resonance in studies of strongly correlated electrons \cite{Hartnoll:2021ydi}. The relaxational and dissipational dynamics of the quantum-critical fluid is controlled only by the Planckian time $\hbar/k_B T$ \cite{CHN89,SY}, and is entirely independent of the energy scale of the interactions between the microscopic degrees of freedom. This appears remarkable when viewed from the lens of the Landau theory of quasiparticles, where the relaxational dynamics is determined by collisions between quasiparticles, and the collision rate is proportional to the square of the interaction strength via Fermi's Golden Rule. 

A fresh perspective on the dynamics of systems without quasiparticles has recently appeared in form of out-of-time-order correlators (OTOCs) \cite{bound}. These relate the absence of quasiparticles to the rapid growth of quantum many-body chaos. The Wilson-Fisher fixed point is also expected to become chaotic in a time of order $\hbar / (k_B T)$ \cite{Chowdhury17}, and such a time is the shortest possible \cite{QPTbook}, as has been proven for a class of models with many local degrees of freedom \cite{bound}.

This article will describe perhaps the simplest system without quasiparticle excitations: the Sachdev-Ye-Kitaev (SYK) model of fermions with random interactions \cite{SY93,kitaev_talk,Sachdev15}. The simplicity has allowed proof of the absence of quasiparticles, and the computation of numerous properties of non-quasiparticle dynamics that are intractable for the Wilson-Fisher theory. Moreover, the SYK model provides a realization of the low energy dynamics of black holes with a near-horizon AdS$_2$ geometry \cite{SS10}, and this has led to new insights on the nature of the Bekenstein-Hawking entropy of black holes which resolve long-standing open questions \cite{Preskill91,Maldacena:1997ih}. The SYK model has also led to significant advances in the theory of strange metals in correlated electron compounds \cite{SYKRMP,Patel:2022gdh,CLi24}. Only the black hole connection is described in this chapter, and the connection to strange metals is described in a companion article \cite{SSORE}.
The presentation is designed for condensed matter theorists, in alignment with Michael Fisher's vision of crossing inter-disciplinary boundaries. 

The SYK model has no spatial structure, and should be considered as a critical quantum system in spatial dimension $d=0$. 
Naively extending the results (\ref{e1},\ref{e2}) to $d=0$, we conclude that the $T=0$ entropy of the SYK model should be a universal constant. 
This naive expectation is essentially correct, but we do have to carefully consider the orders of limits of $N \rightarrow \infty$ (where $N$ is the number of fermion flavors) and $T \rightarrow 0$, and the influence of irrelevant operators. For a SYK model with fermion density $\mathcal{Q}$ ($0 \leq \mathcal{Q} \leq 1$), we have for the total entropy $S$ in the thermodynamic limit $N \rightarrow \infty$ followed by $T \rightarrow 0$
\begin{equation}
\frac{S(T, \mathcal{Q})}{k_B} = N(s_0 + \gamma \, k_B T) - \frac{3}{2}\ln \left(\frac{U}{k_B T} \right) - \frac{\ln N}{2} + \mathcal{O}(1/N) \,. \label{SSYK}
\end{equation}

The leading result in (\ref{SSYK}) is the $Ns_0$ term \cite{GPS2}, and this is the analog of the scaling result in (\ref{e2}) in $d=0$. Here $s_0$ is a universal function of the charge $\mathcal{Q}$; for half-filling with $\mathcal{Q}=1/2$, we have 
\begin{align}
s_0 = \frac{\mathcal{G}}{\pi} + \frac{\ln(2)}{4} = 0.4648476991708051 \ldots, 
\end{align}
where $\mathcal{G}$ is Catalan's constant.
So this term indicates a non-zero entropy density in the limit $T \rightarrow 0$. However, this term does not imply an exponentially large ground state degeneracy; determining the ground state degeneracy requires consideration of the limit $T \rightarrow 0$ followed by $N \rightarrow \infty$: there is vanishing entropy density in such a limit, and we will describe the structure of this limit in Section~\ref{sec:SYK}. 
Indeed, the SYK model is the first model to display a non-zero entropy density in the thermodynamic zero temperature limit {\it without} an exponentially large ground state degeneracy. And this remarkable property is one of the keys to the black hole connection \cite{SS10}, as we describe below.

The first correction to $N s_0$ in (\ref{SSYK}) is the term proportional to $\gamma \sim 1/U$, where $U$ is the root-mean-square strength of the interactions. This correction is due to the leading irrelevant operator, associated with the `Schwarzian', 
which will play a crucial role in the analysis in Section~\ref{sec:SYK} \cite{Cotler16,Bagrets17,Maldacena_syk,kitaevsuh,StanfordWitten}. 

Finite $N$ fluctuation corrections have also been computed, and these require an exact path integral over the effective action of the Schwarzian operator. Such a path integral leads to the $(3/2) \ln (T)$ correction in (\ref{SSYK}). The key property of this correction is that it diverges as $T \rightarrow 0$, indicating a breakdown of the $1/N$ expansion in this limit. Such a breakdown is a characteristic feature of `{\it dangerously irrelevant\/}' operators, a concept introduced by Michael Fisher \cite{Gunton1973}. By comparing the $\gamma$ term with the $(3/2) \ln (T)$, we see that the latter becomes larger at a temperature scale $T \sim e^{-N}$, and so the Schwarzian operator is dangerously irrelevant below this exponentially small $T$. Section~\ref{sec:SYK} will discuss the $\ln N$ term \cite{kitaevsuh,GKST} in (\ref{SSYK}), and the nature of the spectrum of the SYK model at low energy scales.

Let us now turn to the quantum theory of charged black holes. There are a variety of reasons for suspecting that there is a connection between this seemingly unrelated system and the SYK model \cite{SS10}:
\begin{itemize}
\item It has long been known that black holes exhibit `Planckian' dynamics \cite{cvc}; more specifically, the ring-down time $\tau_r$ of quasi-normal modes of black holes, when expressed in terms of the black hole Hawking temperature $T$ \cite{Hawking74}, is of the order of $\hbar/(k_B T)$.
\item The AdS/CFT correspondence of string theory \cite{Maldacena} describes a duality between conformal field theories in $d+1$ spacetime dimensions and quantum gravity in AdS$_{d+2}$ space for $d \geq 1$. For the $d=0$ case, it was pointed out \cite{SS10} that conformal structure of correlators in SYK model was identical to correlators in the AdS$_2$ horizon of charged black holes \cite{Faulkner09}.
\item Charged black holes have a non-zero entropy as $T \rightarrow 0$ at fixed black hole charge $Q$. This matches the behavior of the large $N$ limit of the SYK model in (\ref{SSYK}). Hawking's famous result for black hole entropy is \cite{Hawking74}
\begin{align}
S = \frac{\mathcal{A} (T) c^3}{4 \hbar G}\,,
\end{align}
where $\mathcal{A} (T)$ is the area of the horizon at a temperature $T$. For the case of a black hole with total charge $Q$ in a 3+1 dimensional spacetime which is asymptotically Minkowski, $\mathcal{A}_0 \equiv \mathcal{A} (T \rightarrow 0) = 2 G Q^2/c^4$ is non-zero, and Hawking's result for the low $T$ entropy is \cite{Myers99a,Faulkner09,Sachdev19}
 \begin{equation}
\frac{S(T, Q)}{k_B} = \frac{c^3}{4 \hbar G}\left(\mathcal{A}_0 + 2\sqrt{\pi} \mathcal{A}_0^{3/2} \frac{ k_B T}{\hbar c}  + \mathcal{O}(T^2) \right) \,. \label{SG}
\end{equation}
Note the perfect match of (\ref{SG}) with the $T$ dependence of the first 2 terms of the SYK model in (\ref{SSYK}). The match between $s_0$ term in (\ref{SSYK}) and the $\mathcal{A}_0$ term in (\ref{SG}) was an important factor in the first proposal \cite{SS10} of the connection between charged black holes and the SYK model.  As we will review in Section~\ref{sec:bh}, this match extends also to finite $N$ corrections, which leads to a common many-body density of states. Here, we are viewing (\ref{SG}) as obtained in a small $G$ expansion, corresponding to the large $N$ expansion in (\ref{SSYK}).
\end{itemize}

Hawking's result for black hole entropy \cite{Hawking74} was revolutionary, and raised many questions. Foremost among them was that it did not identify any quantum degrees of freedom whose many-body density of states, $D(E)$, was the exponential of the entropy. The first identification \cite{Strominger96} of possible degrees of freedom appeared using string theory for charged black holes which had supersymmetry at the lowest energies. This theory had an exponentially large degeneracy of stringy states in the zero energy ground state, and it was established that 
\begin{align}
D(E) = e^{S(T \rightarrow 0, Q)/k_B} \delta (E) + \ldots \label{SV}
\end{align}
for supersymmetric black holes, but the form of $D(E)$ for non-zero $E$ was not described. 

\begin{figure}
\begin{center}
\includegraphics[width=6in]{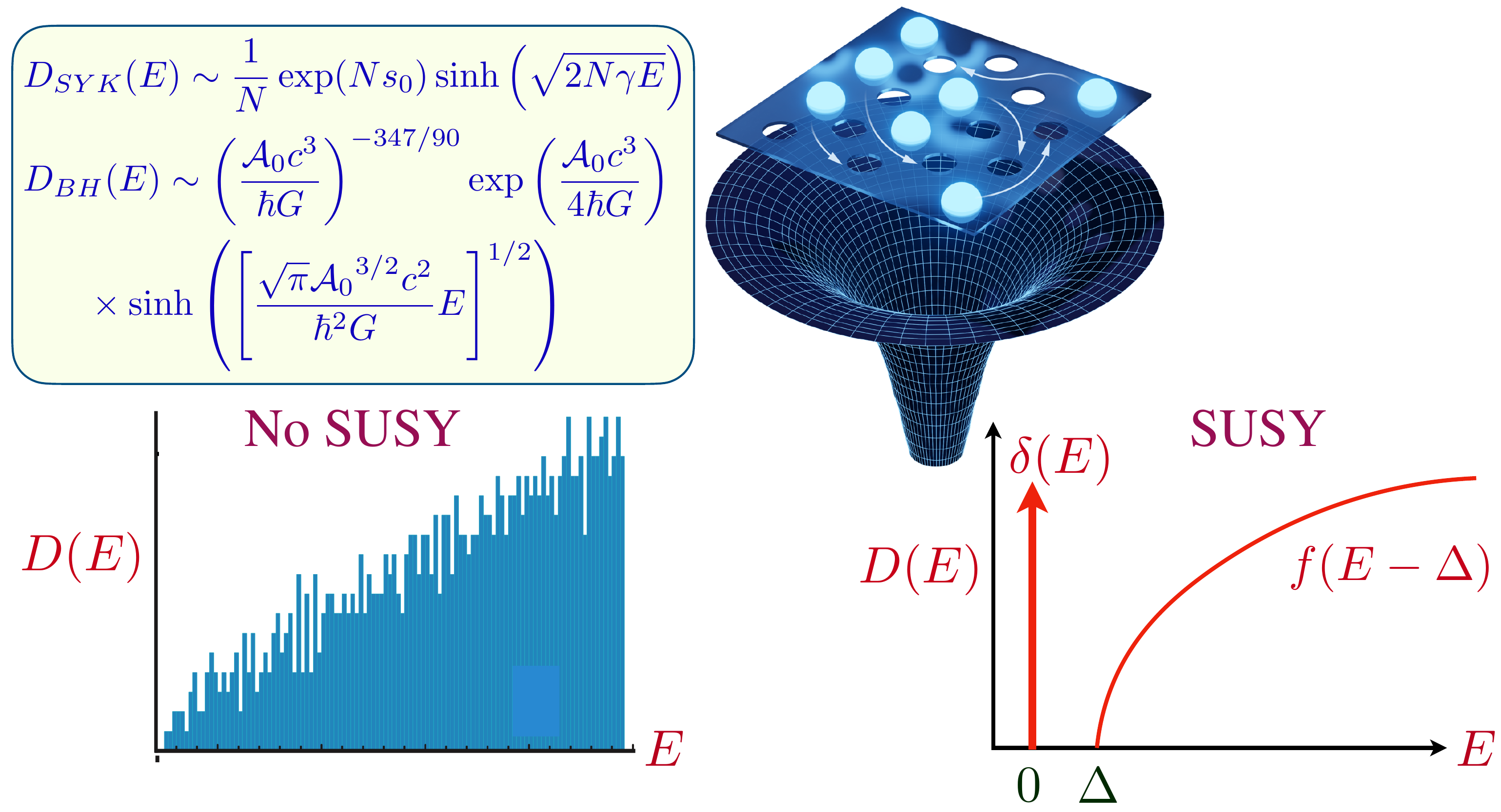}
\end{center}
\caption{Comparison of the many-body densities of states of SYK models ($D_{SYK}(E)$) and charged black holes ($D_{BH}(E)$) with and without supersymmetry (SUSY).
$D_{BH} (E)$ is for a charged black holes in 3+1 dimensional asymptotically Minkowski space without SUSY \cite{Iliesiu:2022onk}, where $\mathcal{A}_0$ is the area of the horizon at $T=0$, see (\ref{DEF}); $D_{SYK} (E)$ is from (\ref{de}) \cite{GKST}.  The left plot is for Majorana SYK, as in Fig.~\ref{fig:grisha}. Both black holes and SYK models with sufficient low energy SUSY have an energy gap $\Delta$, above a delta function with a co-efficient as in (\ref{SV}) \cite{Fu16,StanfordWitten,Heydeman:2020hhw,Iliesiu:2022kny}. Black holes and SYK models without SUSY do not have delta function, nor a gap, but an exponentially dense spacing of levels down to $E=0$; their $D(E)$ vanishes as $E \rightarrow 0$, but with an exponentially large pre-factor, and the latter is sufficient to yield a zero temperature entropy.}
\label{fig:bhdos}
\end{figure}
Here we will review work building on the SYK model which has obtained the form of $D(E)$ for a generic charged black hole without 
supersymmetry; see Fig.~\ref{fig:bhdos}, which summarizes the main results. 
The $D(E)$ of such black holes does {\it not\/} have an exponentially large degeneracy of the ground state. Indeed, there are no large degeneracies (apart from order unity degeneracies required by symmetries), but the energy levels have an exponentially small spacing. Supersymmetric generalizations of the SYK model have also been studied \cite{Fu16,Heydeman:2022lse}, and these do have the a $D(E)$ of the form in (\ref{SV}), confirming that the exponentially large degeneracy is a special feature of low energy supersymmetry. This connection to supersymmetric SYK models has also allowed determination of the non-zero $E$ contributions to the $D(E)$ of supersymmetric black holes in  (\ref{SV}), and such computations show that there is a gap in the many-body density of states above the zero energy delta function \cite{StanfordWitten,Heydeman:2020hhw,Iliesiu:2022kny}. 

A generic thermodynamic system with an extensive entropy has an energy level spacing which is exponentially small in its size ($N$) at an extensive energy above the ground state. Here, we are interested in the behavior of non-supersymmetric systems as the energy density becomes sub-extensive. For a conventional quantum system, the low energy excitations are quasiparticles, and hence the lowest energy level spacing is of order $1/N^a$ (for some positive constant $a$), implying $D(E) \sim N^a$ at small $E$. In this perspective, the remarkable feature of the SYK model, and of charged black holes, is that the energy level spacing remains of order $e^{-N}$ down to very low energies, and this is linked to the non-zero $T \rightarrow 0$ limits of the first two terms in (\ref{SSYK}) and (\ref{SV}).
Indeed, this exponentially small energy level spacing establishes the absence of quasiparticles (although the converse need not be true): the number of quasiparticle states is $\sim N$, and it is not possible to combine them to obtain an exponentially large number of states with a non-extensive energy. The wavefunctions of the low energy states change completely between successive states which happen to have energies exponentially close to each other, a signal of their chaotic nature. In contrast, in quasiparticle systems, successive states are very similar to each other, and differ only by the motion of a few quasiparticles. 

In Section~\ref{sec:SYK}, we will further describe the finite $N$ corrections in (\ref{SSYK}), and determine their consequences for $D(E)$ at the lowest energies. Remarkably, similar corrections apply also to the black hole result in (\ref{SG}), and lead to similar results for the $D(E)$ of generic charged black holes, which will be presented in Section~\ref{sec:bh}. Section~\ref{sec:wf} will present brief discussion of the $D(E)$ of the Wilson-Fisher theory placed on a sphere, and discuss its relationship to the SYK model and black holes.

\section{The SYK model}
\label{sec:SYK}

\begin{figure}
\begin{center}
\includegraphics[width=2.5in]{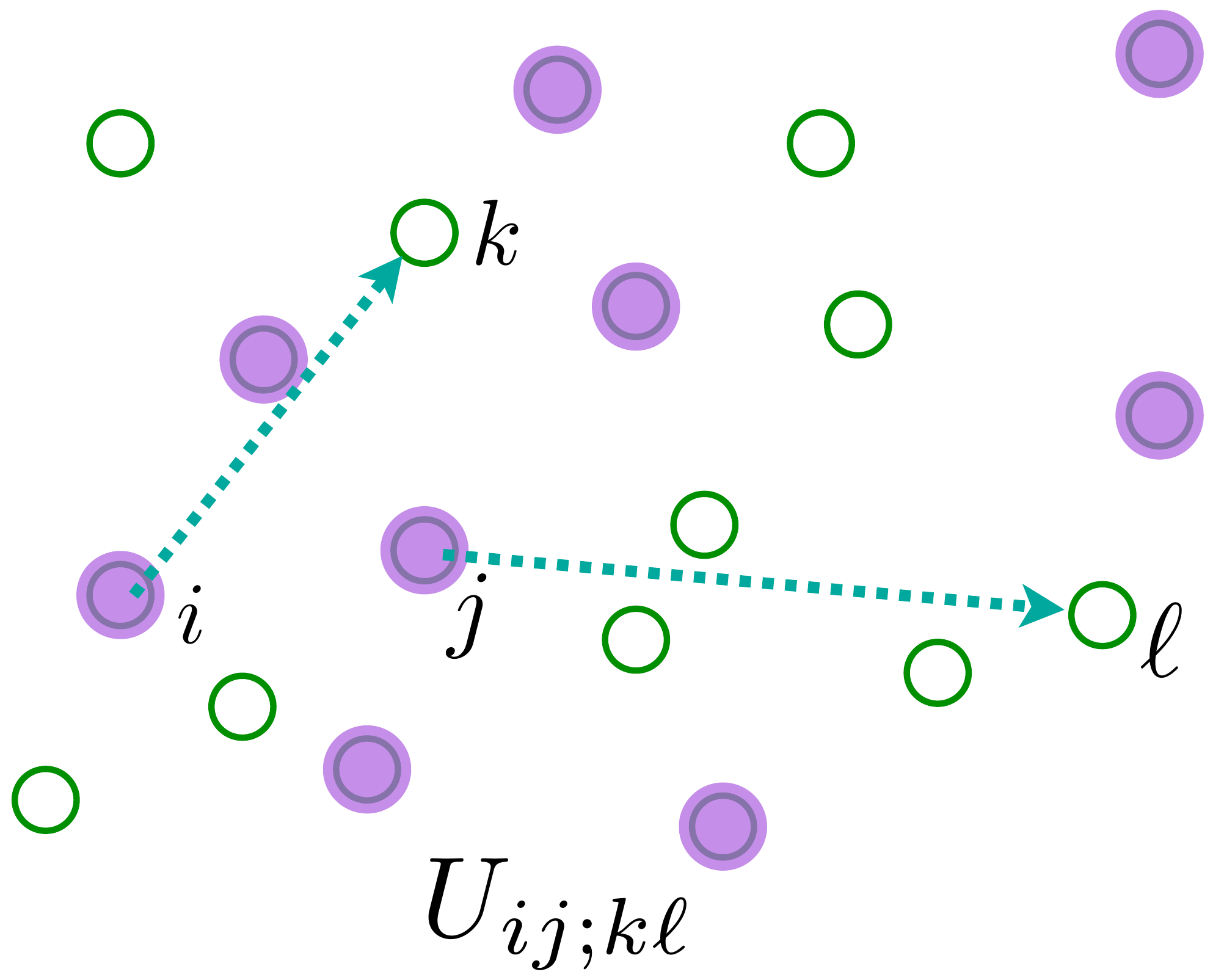}
\end{center}
\caption{The SYK model: fermions undergo the transition (`collision') shown with quantum amplitude $U_{ij;k\ell}$.}
\label{fig:sykU}
\end{figure}

The Hamiltonian of a version of a SYK model is illustrated in Fig.~\ref{fig:sykU}. We take a system with fermions $c_i$, $i=1\ldots N$ states. Depending upon physical realizations, the label $i$ could be position or an orbital, and it is best to just think of it as an abstract label of a fermionic qubit with the two states $\left|0 \right\rangle$ and $c_i^\dagger \left|0 \right\rangle$. We now place $\mathcal{Q} N$ fermions in these states, so that a density $\mathcal{Q} \approx 1/2$ is occupied, as shown in Fig.~\ref{fig:sykU}. 
The quantum dynamics is restricted to {\it only\/} have a `collision' term between the fermions, analogous to the right-hand-side of the Boltzmann equation. However, in stark contrast to the Boltzmann equation, we will not make the assumption of statistically independent collisions, and will account for quantum interference between successive collisions: this is the key to building up a many-body state with non-trivial entanglement. So a collision in which fermions move from sites $i$ and $j$ to sites $k$ and $\ell$ is characterized not by a probability, but by a quantum amplitude $U_{ij;k\ell}$, which is a complex number.

The model so defined has a Hilbert space of order $2^N$ states, and a Hamiltonian determined by order $N^4$ numbers $U_{ij;k\ell}$. Determining the spectrum or dynamics of such a Hamiltonian for large $N$ seems like an impossibly formidable task. But if we now make the assumption that the $U_{ij;k\ell}$ are statistically independent random numbers, remarkable progress is possible. Note that we are not considering an ensemble of SYK models with different $U_{ij;k\ell}$, but a single fixed set of $U_{ij;k\ell}$. Most physical properties of this model are self-averaging at large $N$, and so as a technical tool, we can rapidly obtain them by computations on an ensemble of random $U_{ij;k\ell}$. In any case, the analytic results we now describe have been checked by numerical computations on a computer for a fixed set of $U_{ij;k\ell}$.
We recall that even for the Boltzmann equation, there was an ensemble average over the initial positions and momenta of the molecules that was implicitly performed.

Specifically, the Hamiltonian in a chemical potential $\mu$ is \cite{SY93,kitaev_talk,Sachdev15}
\begin{align}
&\mathcal{H} = \frac{1}{(2 N)^{3/2}} \sum_{i,j,k,\ell=1}^N U_{ij;k\ell} \, c_i^\dagger c_j^\dagger c_k^{\vphantom \dagger} c_\ell^{\vphantom \dagger} 
-\mu \sum_{i} c_i^\dagger c_i^{\vphantom \dagger} \label{HH} \\
& ~~~~~~c_i c_j + c_j c_i = 0 \quad, \quad c_i^{\vphantom \dagger} c_j^\dagger + c_j^\dagger c_i^{\vphantom \dagger} = \delta_{ij}\\
&~~~~\mathcal{Q} = \frac{1}{N} \sum_i c_i^\dagger c_i^{\vphantom \dagger} \, ; \quad
[\mathcal{H}, \mathcal{Q}] = 0\, ; \quad  0 \leq \mathcal{Q} \leq 1\,,
\end{align}
and its large $N$ limit is most simply taken graphically, order-by-order in $U_{ij;k\ell}$, and averaging over $U_{ij;k\ell}$ as independent random variables with $\overline{U_{ij;k\ell}} = 0$ and $\overline{|U_{ij;k\ell}|^2} = U^2$.
\begin{figure}
\begin{center}
\includegraphics[width=2.5in]{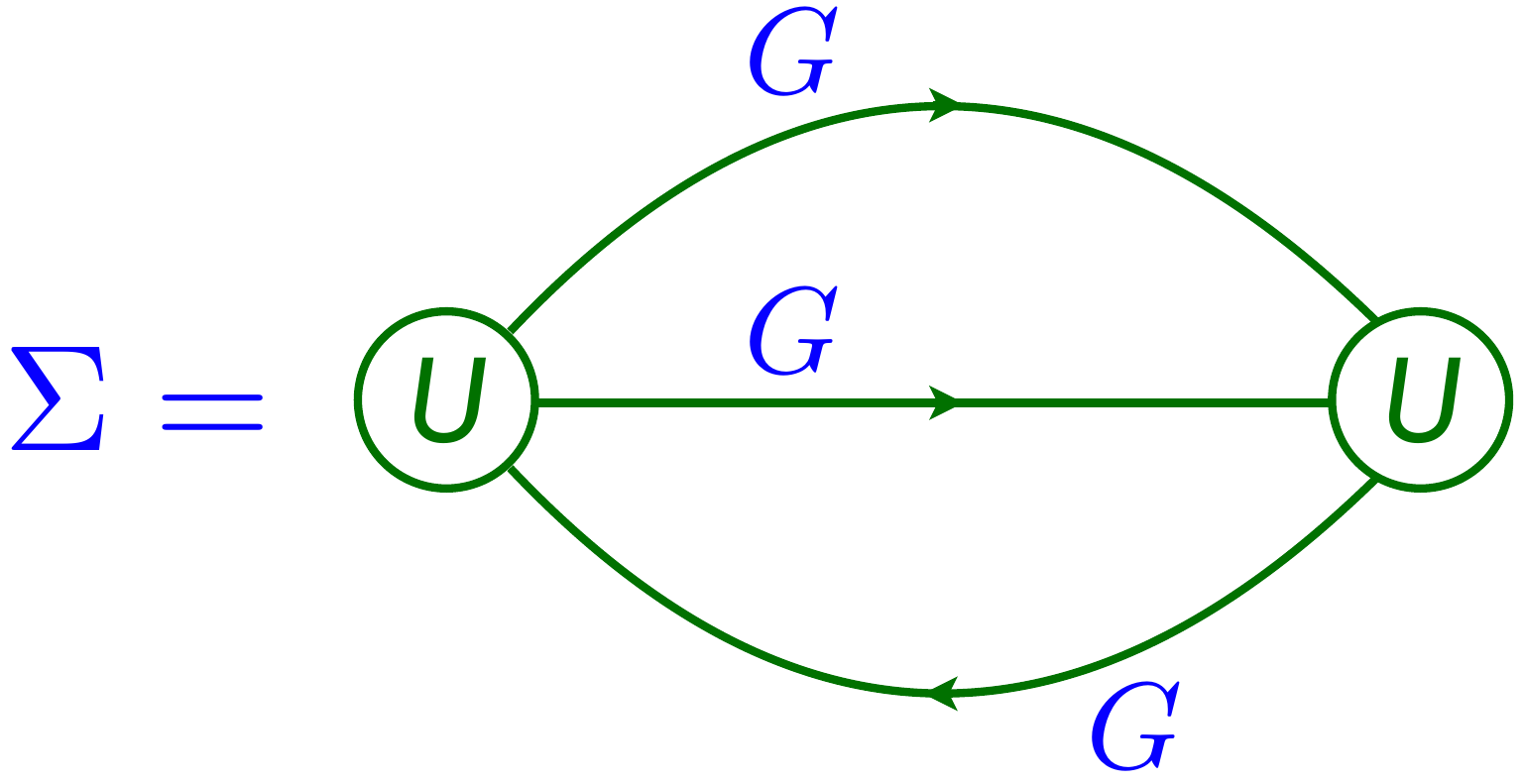}
\end{center}
\caption{Self-energy for the fermions of $\mathcal{H}$ in (\ref{HH}) in the limit of large $N$. The intermediate Green's functions are fully renormalized.}
\label{fig:sygraph}
\end{figure}
In the large $N$ limit, only the graph for the self energy in Fig.~\ref{fig:sygraph} survives, and the on-site fermion Green's function is given by the solution of the following equations
\begin{align}
G(i\omega) = \frac{1}{i \omega + \mu - \Sigma (i\omega)} \quad &, \quad \Sigma (\tau) = -  U^2 G^2 (\tau) G(-\tau) \nonumber \\
G(\tau = 0^-) & = \mathcal{Q}\,, \label{sy1}
\end{align}
where $\tau$ is imaginary time, and $\omega$ is imaginary frequency. 

For general $\omega$ and $T$, Eqs.~(\ref{sy1}) have to be solved numerically. But an exact analytic solution is possible in the limit $\omega, T \ll U$. 
At $T=0$, the asymptotic forms can be obtained straightforwardly \cite{SY93}
\begin{align}
G(i \omega) \sim -i \mbox{sgn} (\omega) |\omega|^{-1/2} \quad, \quad \Sigma(i \omega) - \Sigma (0) \sim -i \mbox{sgn} (\omega) |\omega|^{1/2}\,,
\label{sy10}
\end{align}
and a more complete analysis allows one to obtain the exact form at non-zero $T$ ($\hbar = k_B = 1$) \cite{Parcollet1}
\begin{align}
G (\omega)  = \frac{-i C e^{-i \theta}}{(2 \pi T)^{1/2}}
\frac{\Gamma \left( \displaystyle \frac{1}{4} - \frac{i  \omega}{2 \pi T} + i \mathcal{E} \right)}
{\Gamma \left(  \displaystyle \frac{3}{4} - \frac{i \omega }{2 \pi T} + i \mathcal{E} \right)} \quad\quad  |\omega|, T \ll U \,. \label{sy2}
\end{align}
Here is $\mathcal{E}$ is a dimensionless number which characterizes the particle-hole asymmetry of the spectral function; both $\mathcal{E}$ and the pre-factor $C$ are determined by an angle $-\pi/4 < \theta < \pi/4$
\begin{align}
e^{2 \pi \mathcal{E}} = \frac{\sin(\pi/4 + \theta)}{\sin(\pi/4 - \theta)} \quad, \quad  C = \left( \frac{\pi}{U^2 \cos (2 \theta) }\right)^{1/4}\,,
\end{align}
and the value of $\theta$ is determined by a Luttinger relation to the density $\mathcal{Q}$
\begin{align}
\mathcal{Q} = \frac{1}{2} - \frac{\theta}{\pi} - \frac{\sin(2 \theta)}{4}\,.
\end{align}

The striking property of (\ref{sy2}) is its conformally invariant form, which shows that at low energies the characteristic frequency scale is determined solely by the Planckian frequency scale $k_B T/\hbar$ and is independent of the value of $U$, as discussed in Section~\ref{sec:intro}.

\subsection{$G$-$\Sigma$ theory}

We analyze fluctuations about the large $N$ solution (\ref{sy2}) by the path integral method \cite{Maldacena_syk,kitaevsuh}. 

After introducing replicas $a=1\dots n$, and integrating out the disorder, the Grassman path integral of the Hamiltonian  $\mathcal{H}$ in (\ref{HH}) can be written as 
\begin{align}
\mathcal{Z} &= \int \mathcal{D} c_{ia}(\tau) \exp \left[ - \sum_{ia} \int_0^\beta d\tau \, c_{ia}^\dagger
\left( \frac{\partial}{\partial \tau} - \mu \right) c_{ia} \right. \nonumber \\
& ~~~~~~~~~\left. - \frac{U^2}{4 N^3} \sum_{ab} \int_0^\beta 
d\tau d \tau' \left| \sum_i c_{i a}^\dagger (\tau) c_{i b} (\tau') \right|^4 \right]\,. \label{ZZ}
\end{align}
For simplicity, we neglect the replica indices, and introduce the identity
\begin{align}
1 &= \int \mathcal{D} G(\tau_1, \tau_2) \mathcal{D} \Sigma(\tau_1, \tau_2) \exp \Biggl[ - N \int_0^\beta d\tau_1 d\tau_2 \Sigma(\tau_1, \tau_2) \Biggl( 
G(\tau_2, \tau_1) \nonumber \\
&~~~~~~~~~~~~~~~~~~~~~~~~~~~~~~~~~~~~+ \frac{1}{N} \sum_i c_{i} (\tau_2) c_i^\dagger (\tau_1) \Biggr) \Biggr] \,. 
\end{align}
Then the partition function can be written as 
a path integral with an action $S$ analogous to a Luttinger-Ward functional
\begin{align}
\mathcal{Z} &= \int \mathcal{D} G(\tau_1, \tau_2) \mathcal{D} \Sigma (\tau_1, \tau_2) \exp (-N I) \nonumber \\
I &= \ln \det \left[ \delta(\tau_1 - \tau_2) (\partial_{\tau_1} + \mu) - \Sigma (\tau_1, \tau_2) \right] \nonumber \\
&~~~
+ \int d \tau_1 d \tau_2  \left[ \Sigma(\tau_1, \tau_2) G(\tau_2, \tau_1) + (U^2/2) G^2(\tau_2, \tau_1) G^2(\tau_1, \tau_2) \right] \label{GSigma}
\end{align}
This is a `$G$-$\Sigma$' theory \cite{GPS2,Sachdev15,kitaevsuh,Maldacena_syk}, written as a path integral over bi-local in time functions $G(\tau_1, \tau_2$ and $\Sigma(\tau_1, \tau_2)$, in contrast to the path integral over $N$ fermionic fields in (\ref{ZZ}).
The saddle point of (\ref{GSigma}) yields the equations (\ref{sy1}) for the Green's function $G(\tau_1 - \tau)$,  and the self energy $\Sigma (\tau_1 - \tau_2)$. 

\subsection{Emergent symmetries}

To analyze fluctuations of (\ref{GSigma}) about its saddle point, we need a better understanding of its symmetries, as this will help motivate an effective action for the lowest energy fluctuations.
At frequencies $\ll U$, the time derivative in the determinant is less important, and without it the path integral is invariant under the time reparametrization ($f(\sigma)$) and gauge ($\phi(\sigma)$) transformations
\begin{align}
\tau &= f (\sigma) \nonumber \\
G(\tau_1 , \tau_2) &= \left[ f' (\sigma_1) f' (\sigma_2) \right]^{-1/4}  e^{-i \phi (\sigma_1) + i \phi (\sigma_2)} \, G(\sigma_1, \sigma_2) \nonumber \\
{\Sigma} (\tau_1 , \tau_2) &= \left[ f' (\sigma_1) f' (\sigma_2) \right]^{-3/4} e^{-i \phi (\sigma_1) + i \phi (\sigma_2)} \, {\Sigma} (\sigma_1, \sigma_2) 
\end{align}
where $f(\sigma)$ and $\phi(\sigma)$ are arbitrary functions.

We also need a better understanding of the symmetries of the saddle-point solutions of (\ref{GSigma}), which we denote as $G_s (\tau_1 - \tau_2)$ and $\Sigma_s (\tau_1 - \tau_2)$. The solutions in (\ref{sy10}) written in imaginary time are
\begin{eqnarray}
G_s (\tau_1 - \tau_2) &\sim& (\tau_1-\tau_2)^{-1/2} \nonumber \\ 
\Sigma_s (\tau_1 - \tau_2) &\sim& (\tau_1 - \tau_2)^{-3/2}. \label{sy10a}
\end{eqnarray}
The saddle point will be invariant under a reparamaterization $f(\tau)$ when
choosing $G(\tau_1, \tau_2) = G_s (\tau_1 - \tau_2)$ leads to a transformed
$\widetilde{G}(\sigma_1 , \sigma_2) = G_s (\sigma_1 - \sigma_2)$ (and similarly for $\Sigma$). It turns out this
is true for (\ref{sy10a}) only for the SL(2, R) transformations under which 
\begin{align}
f(\tau) = \frac{a \tau + b}{c \tau + d} \quad, \quad ad -bc =1.
\label{sl2r}
\end{align}
So the (approximate) reparametrization symmetry is spontaneously broken
down to SL(2, R) by the saddle point.

Let us also note the extension of (\ref{sl2r}) to $T>0$. The $T>0$ solution in (\ref{sy2}) is the Fourier transform of 
\begin{align}
G(\tau_1 - \tau_2) = - A \frac{e^{-2 \pi \mathcal{E} T (\tau_1 - \tau_2) }}{\sqrt{1 + e^{-4 \pi \mathcal{E}}}} \left( \frac{T}{\sin(\pi T (\tau_1 - \tau_2))} \right)^{2 \Delta}\,. \label{GT}
\end{align}
This is invariant under PSL(2, R) transformations which map the thermal circle onto itself, and an associated gauge transformation
\begin{align}
\frac{\tan (\pi T f(\tau))}{\pi T}  &=  \frac{a \displaystyle \frac{\tan(\pi T \tau)}{\pi T} + b }{c \displaystyle \frac{\tan(\pi T \tau)}{\pi T} + d } \quad, \quad ad - bc =1, \nonumber \\
~ \nonumber \\
-i \phi (\tau) &= - i \phi_0 + 2 \pi \mathcal{E} T ( \tau - f(\tau))\,. \label{sl2rT}
\end{align}
Indeed, we can derive the $T>0$ form (\ref{GT}) for (\ref{sy10a}) by using (\ref{sl2rT}) \cite{Sachdev15}.

\subsection{From the SYK model to the Schwarzian action}
\label{sec:SYKSchwarz}

We now return to the path integral in (\ref{GSigma}). We focus on the vicinity of the saddle point, $G_s$, $\Sigma_s$, and the low energy  ``Nambu-Goldstone'' modes associated with breaking time reparameterization and U(1) gauge symmetries by writing \cite{Maldacena_syk,kitaevsuh}
\begin{align}
G(\tau_1, \tau_2) = [f'(\tau_1) f'(\tau_2)]^{1/4} G_s (f(\tau_1) - f(\tau_2)) e^{i \phi (\tau_1) - i \phi(\tau_2)}
\end{align}
(and similarly for $\Sigma$). Then the path integral (\ref{GSigma}) is approximated by
\begin{align}
\mathcal{Z} = \int \mathcal{D}f(\tau) \mathcal{D} \phi (\tau) e^{-E_0/T + Ns_0 -N I_{\rm eff} [f, \phi]}\,, \label{ZSch}
\end{align}
where $E_0 \propto N$ is the ground state energy. Demanding that $S_{\rm eff}$ is invariant under the symmetries in (\ref{sl2rT}), we obtain
the following effective action in a gradient expansion (this can be viewed as `non-linear sigma model' associating with the breaking of time reparameterizations to SL(2,R))
\begin{align}
I_{\rm eff} [f, \phi] = \frac{K}{2} \int_0^{1/T} d \tau (\partial_\tau \phi + i (2 \pi \mathcal{E} T) \partial_\tau \epsilon)^2 -  \frac{\gamma}{4 \pi^2} \int_0^{1/T} d \tau \, \Bigl\{ \tan (\pi T (\tau + \epsilon (\tau))), \tau\Bigr\}, \label{Seff}
\end{align}
where $f(\tau) \equiv \tau + \epsilon (\tau)$, the couplings $K$, $\gamma$, and $\mathcal{E}$ can be related to thermodynamic derivatives.
Specifically, the coupling $\gamma$ is the same as that appearing in (\ref{SSYK}). We have used the Schwarzian
\begin{align}
\Bigl\{g, \tau\Bigr\} \equiv \frac{g'''}{g'} - \frac{3}{2} \left( \frac{g''}{g'} \right)^2 \,,
\end{align}
which is obtain from the requirement that effective action at $T=0$ obey
\begin{align}
I_{\rm eff} \left[ f(\tau) = \frac{a \tau + b}{c \tau + d} , \phi (\tau) = 0 \right] = 0 \,.
\end{align}

\subsection{Many-Body Density of States}
\label{sec:dos}

\begin{figure}
\begin{center}
\includegraphics[width=4.5in]{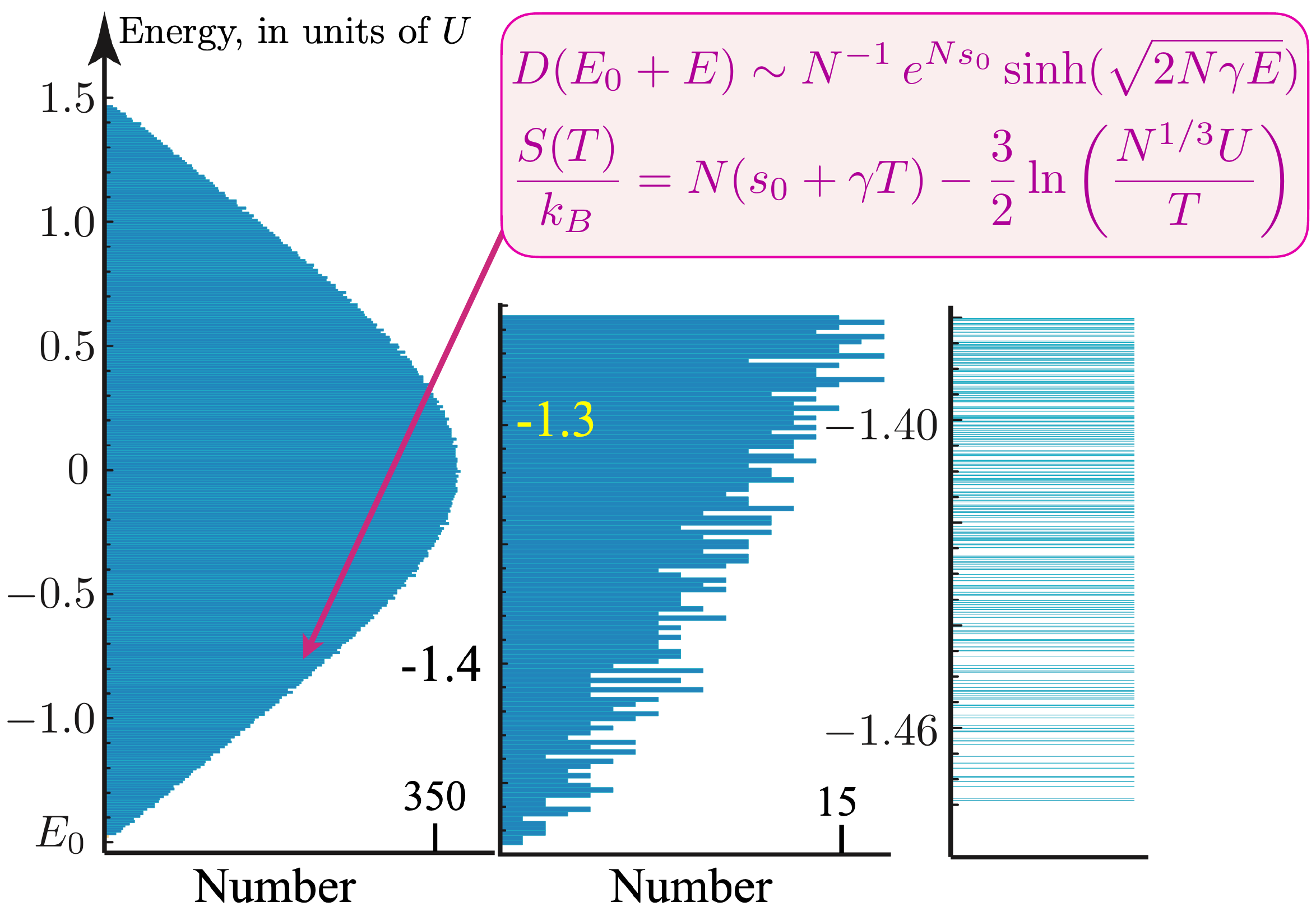}
\end{center}
\caption{Plot of the 65536 many-body eigenvalues of a $N = 32$ Majorana SYK Hamiltonian (credit: G. Tarnopolsky); however, the analytical results quoted here are for the SYK model with complex fermions which has a similar spectrum. The coarse-grained low energy and 
low temperature behavior is described by (\ref{de}) and (\ref{SSYK}). 
}
\label{fig:grisha}
\end{figure}
It now remains to evaluate the path integral in (\ref{ZSch}). Remarkably, this path integral can be evaluated exactly \cite{StanfordWitten}. 
We will not present the details here, and refer the reader to the literature \cite{Cotler16,Bagrets17,Maldacena_syk,kitaevsuh,StanfordWitten,GKST}.
The final result is best expressed in terms of the many-body density of states, related to the partition function by a Laplace transform
\begin{align}
\mathcal{Z} (T, \mathcal{Q}) = \int_{0^-}^{\infty} D(E, \mathcal{Q}) e^{-E/T}\,,\label{Laplace}
\end{align}
while the entropy is given by
\begin{align}
S(T, \mathcal{Q}) = \left. \frac{\partial}{\partial{T}} \left( T \ln \mathcal{Z}(T, \mathcal{Q}) \right) \right|_{\mathcal{Q}}\,. \label{SF}
\end{align}
This procedure yields the main result for the SYK model (see Fig.~\ref{fig:grisha})
\begin{equation}
D(E, \mathcal{Q}) \sim \frac{1}{N} \exp (N s_0) \sinh \left( \sqrt{2 N \gamma E} \right)\,. \label{de}
\end{equation}
It should be noted that the $1/N$ pre-factor does not follow from the low energy theory in (\ref{ZSch}), but requires an analysis of the full partition function in (\ref{GSigma}) \cite{GKST}; this applies also to the $\ln N$ term in (\ref{SSYK}). The remaining terms in (\ref{de}) are more universal, and rely mainly on the SL(2,R) symmetry of (\ref{Seff}).
The result for the entropy in (\ref{SSYK}) now follow from (\ref{Laplace}) and (\ref{de}). 
The thermodynamic limit of the entropy $\lim_{N \rightarrow \infty} S(T)/N$ in (\ref{SSYK}) yields the microcanonical entropy 
\begin{align}
S(E)/k_B = Ns_0 + \sqrt{2N \gamma E}\,, 
\end{align}
and this connects to the extensive $E$ limit of (\ref{de}) after using Boltzmann's formula $D(E) \sim \exp (S(E)/k_B)$.
We summarize these results, and compare with numerics on the SYK model in Fig.~\ref{fig:grisha} and Fig.~\ref{fig:bhdos}.

\section{Charged black holes}
\label{sec:bh}

This section will begin by reviewing Hawking's computation of the entropy of a Schwarzschild black hole with no net charge \cite{Gibbons_Hawking} in Section~\ref{sec:gh}. This is extended to charged black holes in Section~\ref{sec:cbh}, and subsequent subsections discuss reduction of the near-horizon theory of the charged black hole to the Schwarzian theory.

\subsection{Gibbons-Hawking computation of black hole entropy}
\label{sec:gh}

\begin{figure}
\begin{center}
\includegraphics[width=3.5in]{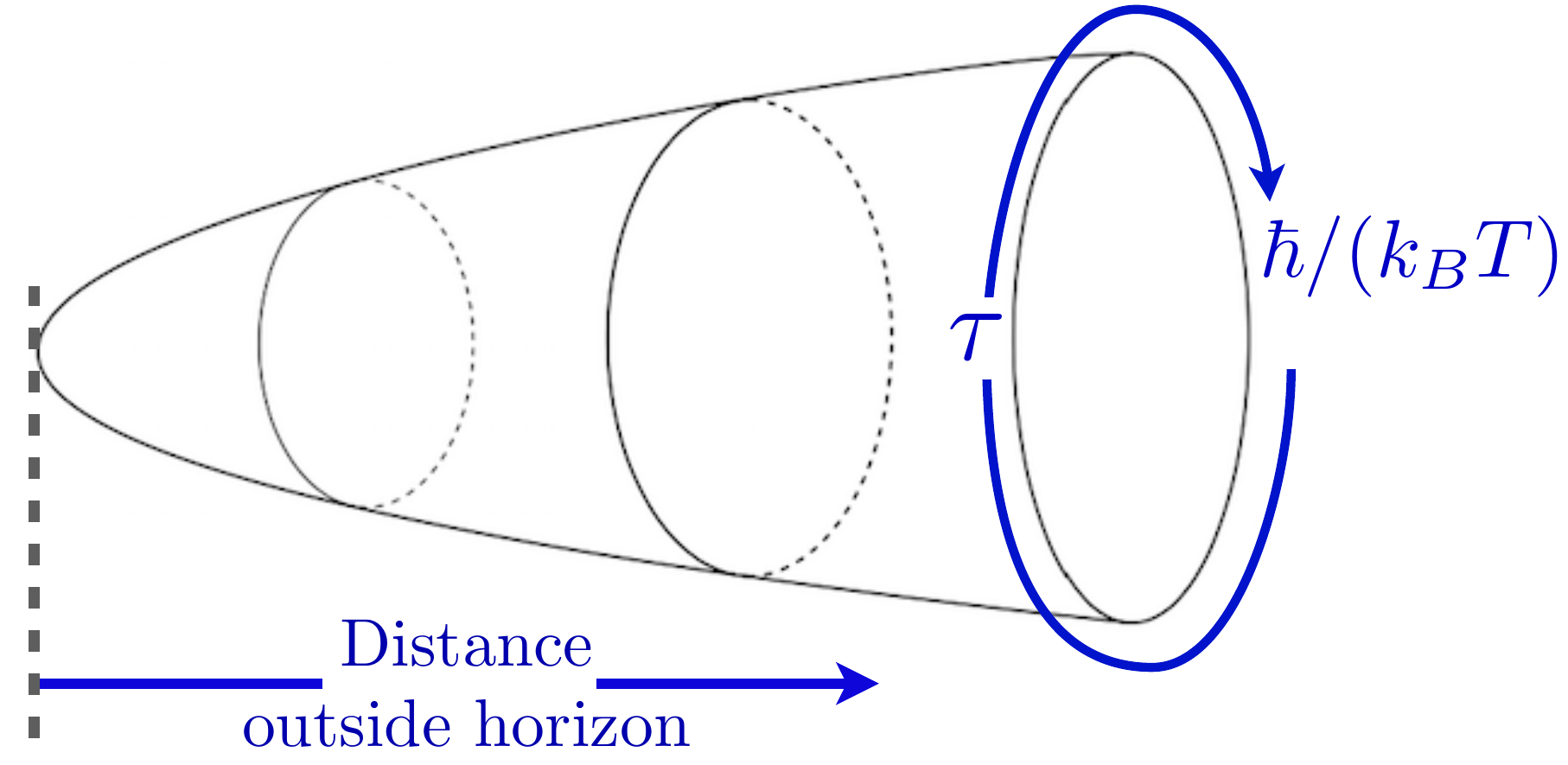}
\end{center}
\caption{Cigar geometry of a Schwarzschild black hole in Euclidean time.
}
\label{fig:cigar}
\end{figure}

\begin{figure}
\begin{center}
\includegraphics[width=3.5in]{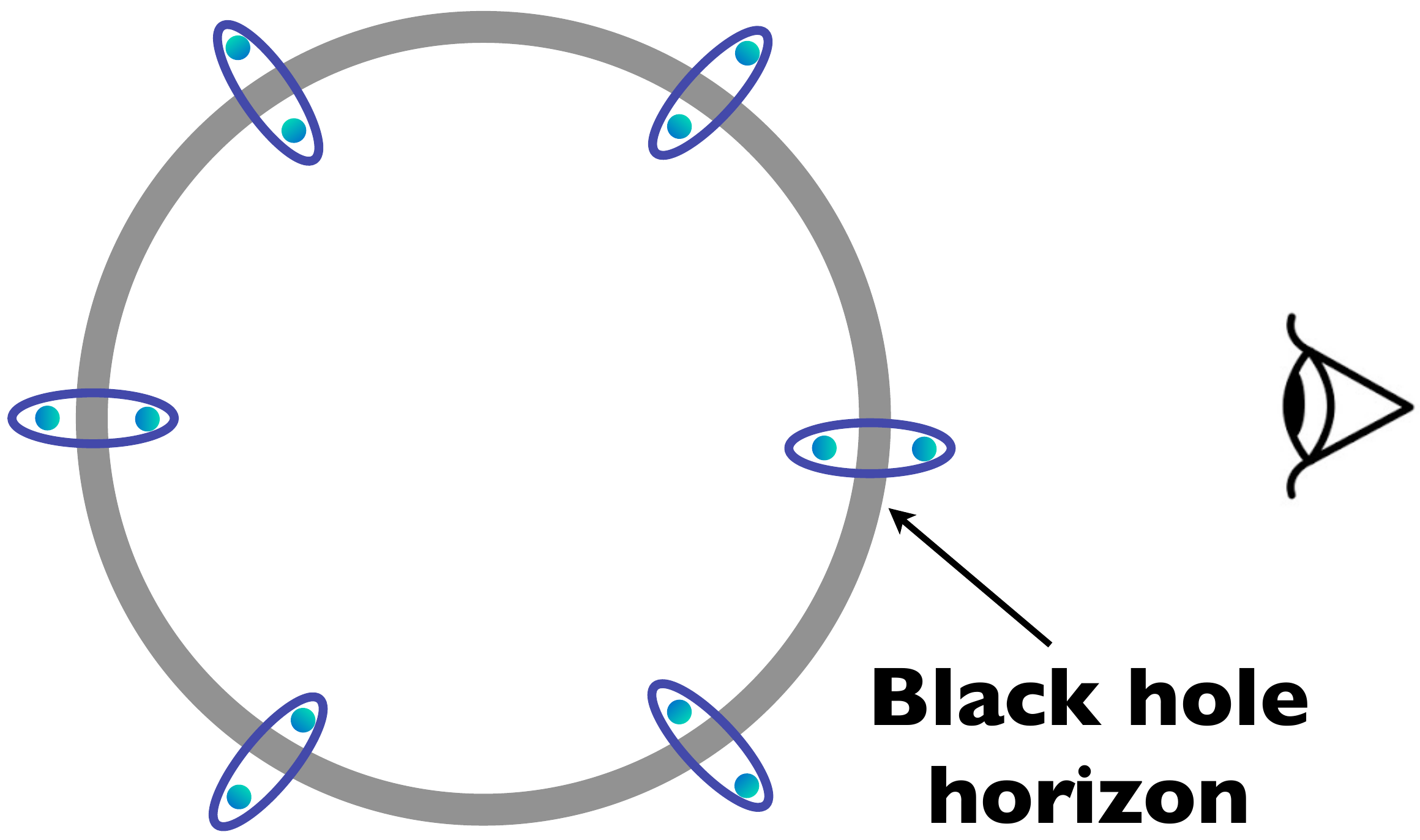}
\end{center}
\caption{Observation of entanglement across the horizon.
}
\label{fig:bhe}
\end{figure}

The Gibbons-Hawking method proceeds by simply evaluating the action of classical Einstein gravity in Euclidean time. Planck's constant appears only via the requirement that the temporal period is $\hbar/(k_B T)$. In Euclidean time, we will see below that the spacetime geometry along the radial and temporal directions is a `cigar' which closes off at the horizon, as shown in Fig.~\ref{fig:cigar}.
Consequently, the computation is performed entirely outside the black hole, and we sidestep our ignorance of the black hole interior. The black hole entropy is then seen to be a consequence of entanglement around the horizon, as shown in Fig.~\ref{fig:bhe}.

The Einstein action for gravity in 3+1 dimensions is
\begin{align}
I_{E} =  \int d^{4} x \sqrt{g} \left[ -\frac{1}{2 \kappa^2} \mathcal{R}_{4} \right] \quad , \quad \mathcal{Z} = \int \mathcal{D} g \exp ( - I_E )\,, \label{IE1}
\end{align}
where $\kappa^2 = 8 \pi G$ is the gravitational constant, $\mathcal{R}_{4}$ is the Ricci scalar.
The Schwarzschild solution of the saddle-point equations is
\begin{align}
ds^2 =  V(r) d\tau^2 + r^2 d \Omega_2^2 + \frac{dr^2}{V(r)}
\end{align}
where $d \Omega_2^2$ is the metric of the $2$-sphere, and
\begin{align}
V(r) = 1  - \frac{m}{r}.
\end{align}
The gravitational mass of the black hole is $M = 2 G m$.
The black hole horizon is at $r=r_0$ where $V(r_0)=0$; so 
\begin{align}
r_0 = m.
\end{align}

The $T>0$ quantum partition function is obtained in a spacetime which is periodic as a function of $\tau$ with period $\hbar/(k_B T)$, as in Fig.~\ref{fig:cigar}. We have to ensure that there is no singularity at the horizon $r_0$ where $V(r_0)=0$. Let us change radial co-ordinates to $y$, where $r=r_0 + y^2$. Then for small $y$
\begin{align}
ds^2 = \frac{4}{V'(r_0)} \left[ \frac{(V'(r_0))^2}{4} \, y^2 d \tau^2 + dy^2 \right] + r_0^2 d \Omega_2^2 = \frac{4}{V'(r_0)} \left[ y^2 d\theta^2 + dy^2 \right] + r_0^2 d \Omega_2^2
\end{align}
The expression in the square brackets is the metric of the flat plane in polar co-ordinates, with radial co-ordinate $y$ and angular co-ordinate $\theta =  V'(r_0) \tau/2$. Smoothness requires periodicity in $\theta$ with period $2 \pi$, and so 
\begin{align}
4 \pi T = V'(r_0) = \frac{1}{m} \,. \label{a4}
\end{align}

The free energy $\beta F = I_E$, where $\beta = 1/T$. So the entropy is (as in (\ref{SF}))
\begin{align}
S = - \frac{\partial F}{\partial T} = \left( \beta \frac{\partial} {\partial \beta} - 1 \right) I_E
\end{align}
However, the metric is $\tau$-independent, and the only explicit dependence of the action is via $I_E = \beta H$. 
Such an action implies $S=0$.

\begin{figure}
\begin{center}
\includegraphics[width=3.5in]{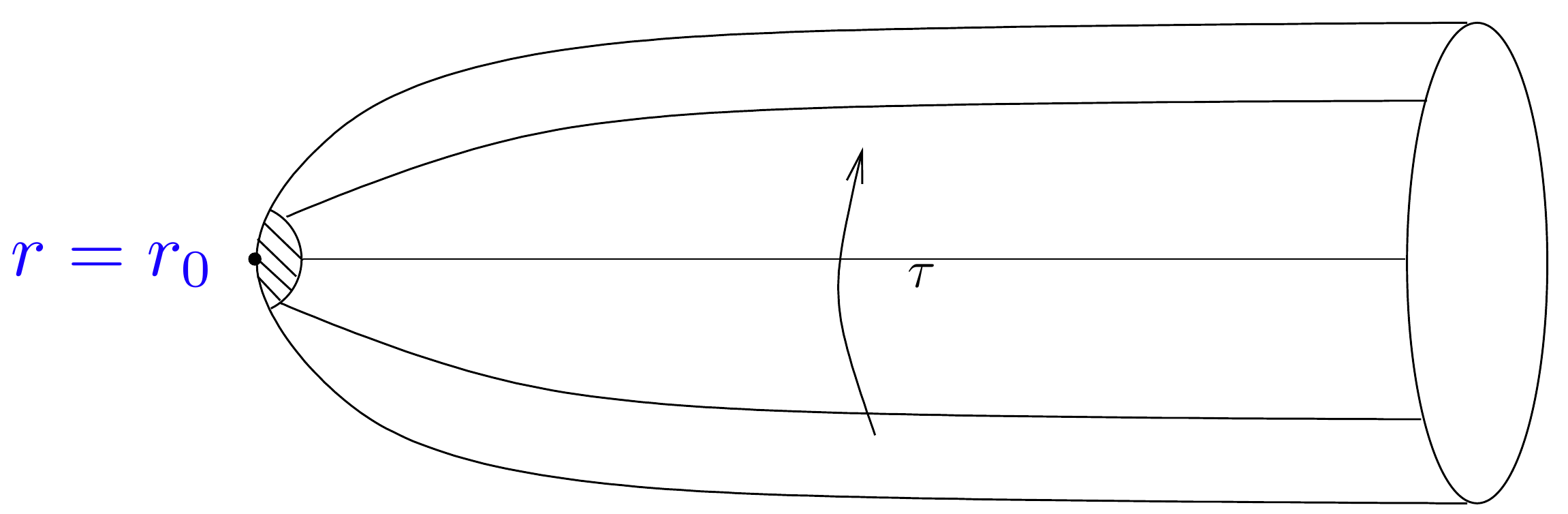}
\end{center}
\caption{The action $I_{\rm grav}$ is evaluated in the shaded region at the end of the cigar (figure adapted from \cite{Ross_review}).
}
\label{fig:cigar2}
\end{figure}

Gibbons and Hawking \cite{Gibbons_Hawking} argued that a proper evaluation of the gravity action requires a boundary term, $I_{GH}$, 
so that the gravity path integral in (\ref{IE1}) is replaced by
\begin{align}
I_{\rm grav} = I_{E} + I_{GH} \quad, \quad I_{GH} =  \int_\partial d^{3} x \sqrt{g_b} \left[ -\frac{1}{\kappa^2} \mathcal{K}_{3} \right] \quad , \quad \mathcal{Z} = \int \mathcal{D} g \exp ( - I_{\rm grav} )\,, \label{IE2}
\end{align}
where $\mathcal{K}_3$ is the extrinsic scalar curvature of the 3-dimensional boundary of spacetime.
$I_{GH}$ is the Gibbons-Hawking boundary term, deduced by the requirement that the Euler-Lagrange equations of $I_{\rm grav}$ co-incide with the Einstein equations, with no additional boundary terms. 
Because the bulk action does not contribute any entropy, we can move the boundary from asymptotically far region to the vicinity
of horizon, as shown in Fig.~\ref{fig:cigar2} \cite{Ross_review}.
The entire contribution to the entropy then comes from $I_{GH}$ evaluated in the vicinity of the co-ordinate singularity at $r=r_0$. 
We evaluate $I_{GH}$ by using the identity
\begin{align}
\int_\partial d^{3} x \sqrt{g_b} \, \mathcal{K}_{3}  =  \frac{\partial}{\partial n} \int_\partial d^{3} x \sqrt{g_b}  
\end{align}
where $n$ is the Gaussian normal co-ordinate of the boundary. Evaluating at $y=\epsilon$, we have 
\begin{align}
\int_\partial d^{3} x \sqrt{g_b} = 2 \pi \epsilon \mathcal{A}
\end{align}
where $\mathcal{A}= 4 \pi r_0^2$ is the area of the horizon.
Combining everything, and noting that the contribution of $I_E$ in the shaded region of Fig.~\ref{fig:cigar2} vanishes in the limit $\epsilon \rightarrow 0$, we have the famous result of Hawking
\begin{align}
S = \frac{ 2\pi\mathcal{A}}{\kappa^2} = \frac{\mathcal{A}}{4 G}.
\end{align}

\subsection{Hawking entropy of a charged black hole}
\label{sec:cbh}

For a charged black hole, we need the Einstein-Maxwell theory of $g$ and a U(1) gauge flux $F = dA$, and simply need to compute the area of the horizon in this modified geometry. The Einstein-Maxwell action is
\begin{align}
I_{EM} =  \int d^{4} x \sqrt{g} \left[ -\frac{1}{2 \kappa^2} \mathcal{R}_{4}  +  \frac{1}{4g_F^2} F^2 \right]\quad , \quad \mathcal{Z}_Q = \int \mathcal{D} g \mathcal{D} A \exp ( - I_{EM} - I_{GH} )\,.
\end{align}
The saddle-point equations now yield a solution with \cite{Myers99a,Faulkner09}
\begin{align}
V(r) & = 1 + \frac{\Theta^2}{r^{2}} - \frac{m}{r} \quad; \quad A_\tau = i \mu \left( 1 - \frac{r_0}{r} \right) \nonumber \\ 
\Theta & =  \frac{\kappa r_0}{\sqrt{2} g_F} \mu \quad ; \quad Q = \frac{4 \pi \mu r_0}{g_F^2} \quad ; \quad S = \frac{2 \pi \mathcal{A}}{\kappa^2}\,,
\end{align}
where $Q$ is the total charge, the chemical potential is $\mu$, and as before the horizon is where $V(r_0) = 0$, the temperature $T = V'(r_0)/(4 \pi)$, and $\mathcal{A} = 4 \pi r_0^2$.
This defines a two parameter family of charged black hole solutions of $I_{EM}$ determined by $T$ and $Q$.

Now we take the limit $T \rightarrow 0$ at fixed $Q$. Then we find the remarkable feature that the horizon radius remains finite
\begin{align}
R_h \equiv r_0 (T \rightarrow 0, Q) = \frac{Q\kappa g_F}{4 \pi} 
\end{align}
In this limit, entropy becomes
\begin{align}
S (T \rightarrow 0, Q) = \frac{4 \pi R_h^2}{G} + \gamma \, T \quad, \quad \gamma \equiv \frac{4 \pi^2 R_h^3}{G}\,, \label{bhentropy}
\end{align}
corresponding to the result in (\ref{SG}).

\begin{figure}
\begin{center}
\includegraphics[width=6in]{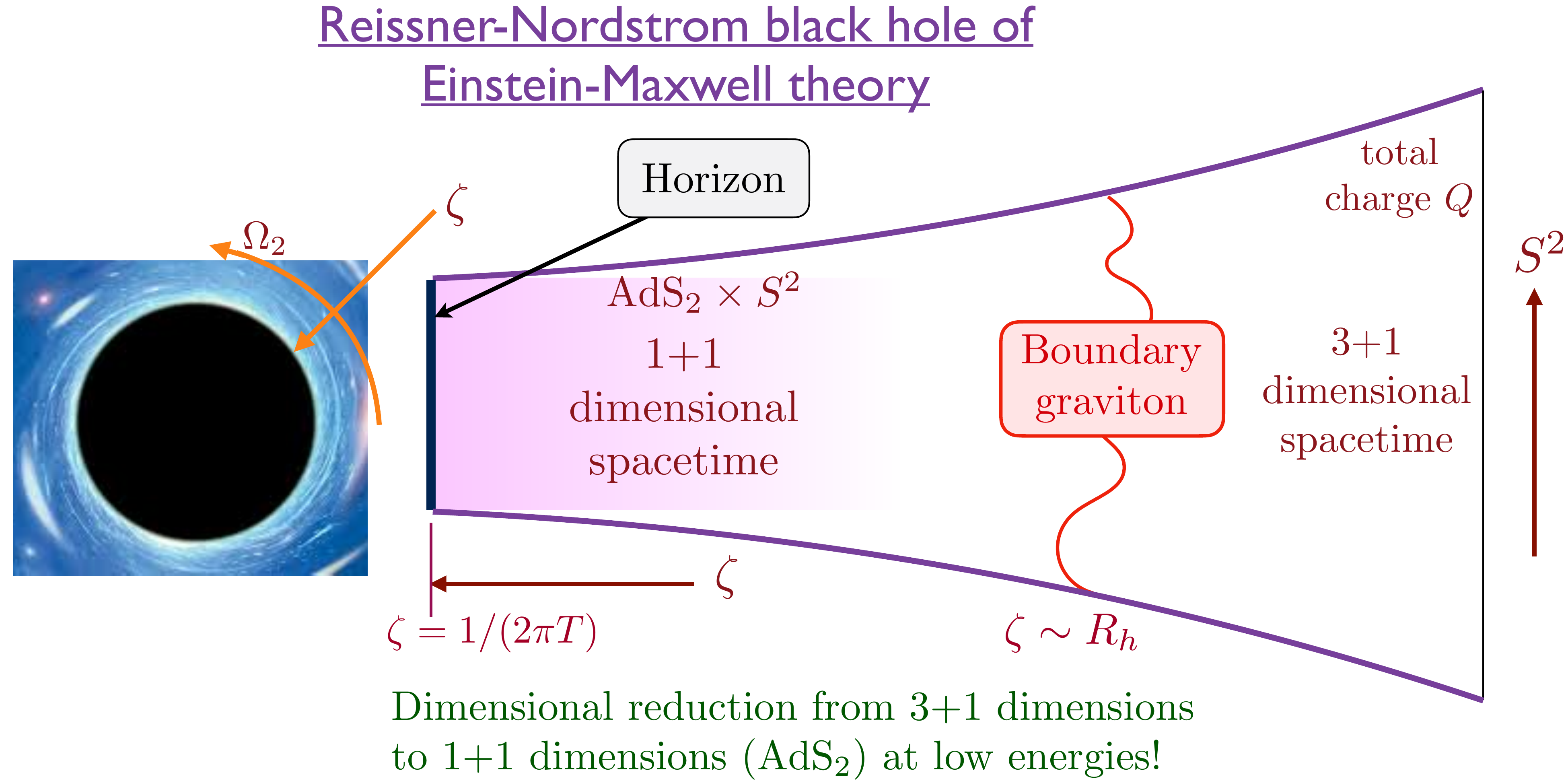}
\end{center}
\caption{Schematic of a charged (Reissner-N\"ordstrom) black hole.
}
\label{fig:rn}
\end{figure}
For the near-horizon metric, it is useful to introduce the co-ordinate $\zeta$
\begin{align}
r = R_h + \frac{R_h^2}{\zeta}
\end{align}
so that the horizon at $T=0$ is at $\zeta = \infty$. Then in the near-horizon regime $R_h \ll \zeta < \infty$ the $T=0$ metric is
\begin{align}
ds^2 = R_h^2 \frac{ d \tau^2 + d \zeta^2}{\zeta^2} + R_h^2 d \Omega_2^2 \label{e10}
\end{align}
The spacetime in (\ref{e10}) is AdS$_2 \times S^2$, as illustrated in Fig.~\ref{fig:rn}. The dominant low energy excitations involve the AdS$_2$ component, and so the near-horizon metric is effectively 1+1 dimensional.

\subsection{AdS$_2$ and its symmetries}
\label{sec:ads2}

The AdS$_2$ metric
\begin{align}
ds^2 = \frac{d \tau^2 + d \zeta^2}{\zeta^2} \label{appads2}
\end{align}
is invariant under isometries which are SL(2,R) transformations. The reader can verify that the co-ordinate change
\begin{align}
\tau' + i \zeta' = \frac{a (\tau + i \zeta) + b}{c (\tau + i \zeta) + d}\,, \quad ad-bc =1\,,
\end{align}
with $a$,$b$,$c$,$d$ real, leaves the AdS$_2$ metric invariant. This is a crucial fact, and establishes a connection to the SL(2,R) invariance of the SYK model in (\ref{sl2r}). 

\begin{figure}
\begin{center}
\includegraphics[width=3.25in]{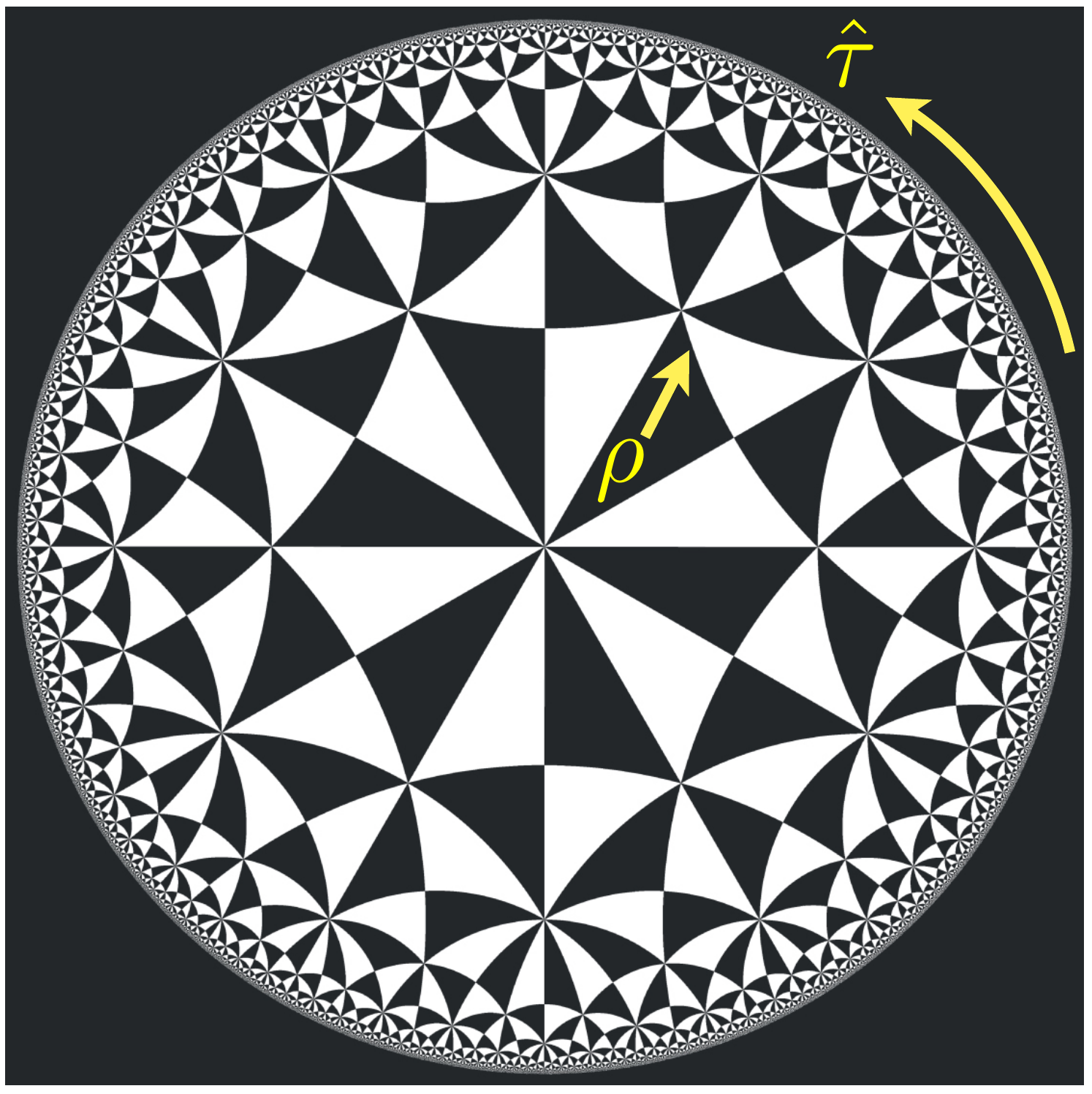}
\end{center}
\caption{AdS$_2$ in $\rho$ and $\hat{\tau}$ co-ordinates, with metric as in (\ref{ads2T}). 
}
\label{fig:ads2}
\end{figure}
As for the SYK model, we can map the $T=0$ metric to $T>0$ metric. The co-ordinate transformation
\begin{align}
\zeta = \frac{1}{\cosh (2 \pi T \rho) - \sinh (2 \pi T \rho) \cos (2 \pi T \hat\tau)} ~, ~ \tau = \frac{\sinh (2 \pi T \rho) \sin (2 \pi T \hat\tau)}{\cosh (2 \pi T \rho) - \sinh (2 \pi T \rho) \cos (2 \pi T \hat\tau)} 
\end{align}
maps the metric to
\begin{align}
ds^2 = 4 \pi^2 T^2 \left[ d \rho^2 + \sinh^2 (2 \pi T \rho) d \hat\tau^2 \right]\,, \label{ads2T}
\end{align}
which is illustrated in Fig.~\ref{fig:ads2}. The $\hat{\tau}$ directions is periodic with period $1/T$.

\subsection{From Einstein-Maxwell theory in 3+1 dimensions to the Schwarzian action}
\label{sec:emsch}

We have shown above how a universal AdS$_2$ metric emerges in the near-horizon region of a charged black hole in 3+1 dimensions (a similar mapping applies in $d+2$ dimensions with $d \geq 2$). This subsection will carry out the same mapping at the level of the action, and so allow us to go beyond the saddle-point approximation at low energies.
 
\subsubsection{From $I_{EM}$ in 3+1 dimensions to $I_{EM,2}$ in 1+1 dimensions.}

This dimensional reduction is carried out the by simply taking all fields dependent only upon the radial co-ordinate $r$ and imaginary time $\tau$. 
We make the metric ansatz \cite{Moitra18}
\begin{align}
ds^2 = \frac{ds_2^2}{\Phi (\zeta, \tau)} + \left[ \Phi (\zeta, \tau) \right]^2 d \Omega_2^2
\end{align}
where $ds_2^2$ is an arbitrary metric in the $(\zeta, \tau)$ spacetime, and $\Phi$ is a scalar field in the $(\zeta, \tau)$ spacetime.
Here we do not discuss the dimensional reduction of the U(1) gauge field $A$, which has been described elsewhere \cite{Moitra18,Sachdev19}.

\subsubsection{JT gravity as the low energy limit of $I_{EM,2}$}

We take the low energy limit of $I_{EM,2}$ by mapping it to a near-horizon theory, $I_{JT}$, in a 1+1 dimensional spacetime with a boundary
\cite{Almheiri:2014cka,Almheiri:2016fws,Maldacena:2016upp,Choi:2021nnq,Choi:2023syx}.
The low energy theory on the $(\zeta, \tau)$ spacetime involves a metric $h$, and a scalar field $\Phi_1$ given by
\begin{align}
\lim_{\zeta \rightarrow \infty} [\Phi(\zeta,\tau)]^2  = R_h^2 + \Phi_1 (\zeta,\tau)\,, 
\end{align}
obeying the action
\begin{align}
I_{JT} = - \frac{2 \pi \mathcal{A}_0}{\kappa^2} + \int d^2 x \sqrt{h} \left[ - \frac{2\pi}{\kappa^2} \, \Phi_1 \left( \mathcal{R}_2 + \frac{2}{R_h^3} \right) \right]   - \frac{4 \pi}{\kappa^2} \int_{\partial} dx \sqrt{h_b} \Phi_1 \, \mathcal{K}_1 \label{jt}
\end{align}
where $\mathcal{A}_0 = 4 \pi R_h^2$ is the area of the horizon at $T=0$, and $\mathcal{K}_1$ is the extrinsic curvature of the one-dimensional boundary $\zeta \rightarrow 0$ where
\begin{align}
h_{\tau\tau} ( \zeta \rightarrow 0) = \frac{R_h^3}{\zeta^2} \quad, \quad \Phi_1 ( \zeta \rightarrow 0) = \frac{2R_h^3}{\zeta} 
\end{align}

\subsubsection{Fluctuations about the AdS$_2$ saddle point of $I_{JT}$}

Einstein gravity in 1+1 dimensions has no graviton, and is `pure gauge'. In the JT-gravity theory with boundary, there is a remnant degree of freedom which is a boundary graviton, as illustrated in Fig.~\ref{fig:rn}. 
The action for this boundary graviton turns out to be the advertised Schwarzian theory: consequently,
the partition function of the $1+1$ dimensional JT gravity theory can be evaluated exactly (here we are ignoring the 
gauge field path integral, which is subdominant at fixed $Q$ \cite{luca20,GKST}).

The partition function of JT gravity is
\begin{align}
\mathcal{Z}_Q = \int \mathcal{D} h \mathcal{D} \Phi_1 \exp \left( - I_{JT} \right)
\end{align}
The action is linear in $\Phi_1$, and the integral over $\Phi_1$ yields a constraint $\mathcal{R}_2 = - 2/R_h^3$ {\it i.e.\/} the metric $h$ is rigidly AdS$_2$.
The only dynamical degree of freedom in JT gravity is a time reparameterization 
along the boundary $\tau \rightarrow f(\tau)$ \cite{Almheiri:2014cka,Maldacena:2016upp}. 
To ensure that the bulk metric obeys its boundary condition, we also have to make the spatial co-ordinate $\zeta$ a function of $\tau$, so we map
$(\tau, \zeta) \rightarrow (f(\tau), \zeta(\tau))$. Then 
the metric obeys its boundary condition 
provided $\zeta (\tau)$ is related to $f(\tau)$ by (here $\zeta_b$ is a small constant whose value cancels in the final result)
\begin{align}
\zeta (\tau) = \zeta_b f' (\tau) + \zeta_b^3  \frac{\left[f'' (\tau) \right]^2}{2 f'(\tau)} + \mathcal{O}(\zeta_b^4 ) 
\end{align}
Finally, we evaluate $I_{GH}$ along this boundary curve.  In this manner we obtain the action
\begin{align}
I_{1,{\rm eff}} [f] = - \frac{2 \pi \mathcal{A}_0}{\kappa^2} -  \frac{\gamma}{4 \pi^2 } \int d \tau \,  \Bigl\{ f(\tau), \tau\Bigr\} \quad, \quad \Bigl\{f, \tau\Bigr\} \equiv \frac{f'''}{f'} - \frac{3}{2} \left( \frac{f''}{f'} \right)^2 
\end{align}
where $\gamma = 32 \pi^3 R_h^3/\kappa^2$ is precisely the linear-$T$ co-efficient in the black hole entropy in (\ref{bhentropy}).

After a conformal map to finite temperature (and ignoring the contribution of the gauge field fluctuation), we can write the low energy partition function of a 3+1-dimensional black hole with charge $Q = 4 \pi R_h/(\kappa g_F)$, as a path integral over a single field $f(\tau)$ in one time dimension:
\begin{align}
\mathcal{Z}_Q = \exp \left( \frac{ 2 \pi \mathcal{A}_0}{\kappa^2} \right) \int \frac{\mathcal{D} f}{|| \mbox{SL(2,R)}||} \exp \left(  \frac{\gamma}{4 \pi^2} \int_0^{1/T} d \tau \,  \Bigl\{ \tan (\pi T f(\tau)), \tau\Bigr\} \right)
\label{ZQQ}
\end{align}
where $\gamma = 32 \pi^3 R_h^3/\kappa^2$, $\mathcal{A}_0 = 4 \pi R_h^2$, and $f (\tau)$ is a monotonic function of $\tau$ obeying
\begin{align}
f (\tau + 1/T) = f(\tau) + 1/T\,.
\end{align}
We divide by the (infinite) volume of the SL(2,R) group because
\begin{align}
\Bigl\{f, \tau \Bigr\} = \Bigl\{ \frac{a f + b}{c f + d} , \tau \Bigr\}
\end{align}
where $a,b,c,d$ are constants with $ad -bc =1$. This is precisely the theory obtained in Section~\ref{sec:SYKSchwarz} for the SYK model. 
The integral over the phase field $\phi$ is not present in (\ref{ZQQ}), but it also appears in the same form after including the dimensional reduction of the gauge field $A$ \cite{Moitra18,Sachdev19}.

\subsection{Many-Body Density of states}
\label{sec:dos2}

With the mapping to the Schwarzian in (\ref{ZQQ}), we can now immediately apply the results of Section~\ref{sec:dos} to the charged black hole. The final results for the entropy and many-body density of states are \cite{Iliesiu:2022onk} 
\begin{align}
\frac{S(T, Q)}{k_B} & = \frac{c^3}{4 \hbar G}\left(\mathcal{A}_0 + 2\sqrt{\pi} \mathcal{A}_0^{3/2} \frac{ k_B T}{\hbar c} \right) - \frac{3}{2}  \ln \left( \frac{(\hbar c^5/G)^{1/2}}{k_B T} \right)  -\frac{559}{180} \ln \left( \frac{\mathcal{A}_0 c^3}{\hbar G} \right) + \ldots\,. \label{SGF} \\
D(E,Q) & \sim \left( \frac{\mathcal{A}_0 c^3}{\hbar G} \right)^{-347/90} \exp\left( \frac{\mathcal{A}_0 c^3}{4 \hbar G} \right) \sinh \left( \left[\sqrt{\pi} \mathcal{A}_0^{3/2} \frac{c^3}{\hbar G} \frac{E}{\hbar c} \right]^{1/2} \right) \,. \label{DEF}
\end{align}
These results should be compared to (\ref{SSYK}) and (\ref{de}) for the SYK model, as in Fig.~\ref{fig:bhdos}. For the entropy, the first two terms are the saddle-point contributions, which are proportional to $1/G$ and $N$ respectively. The $(3/2) \ln T$ terms are identical between (\ref{SGF}) and (\ref{SSYK}), and arise from the Schwarzian path integral. Finally, the $\ln (\mathcal{A}_0)$ term in (\ref{SGF}) is the analog of the $\ln N$ term in (\ref{SSYK}), and these do have different co-efficients, reflecting their sensitivity to microscopics. The $\ln N$ term relies on the fact that (\ref{HH}) contains only 4-fermion terms \cite{kitaevsuh,GKST}, while the $\ln \mathcal{A}_0$ term has a co-efficient dependent upon the number of massless fields in the quantum gravity theory under consideration \cite{Iliesiu:2022onk}. The $\ln \mathcal{A}_0$ term was partly computed in earlier work \cite{Sen12,Banerjee:2023quv}.

The many-body density of states in (\ref{DEF}) is connected to the entropy in (\ref{SGF}) via (\ref{Laplace}), and the $\ln \mathcal{A}_0$ term in the entropy determines the power of $\mathcal{A}_0$ in the pre-factor. However, the $E$ dependence is identical to that in (\ref{de}).

\section{Wilson-Fisher conformal field theory}
\label{sec:wf}

It is instructive to compare the above results for the many-body density of states of the SYK model and charged black holes with those of the Wilson-Fisher conformal field theory. To this end, we place the Wilson Fisher theory in 2+1 dimensions on the two-dimensional surface of a sphere of radius $R$. Via the state-operator correspondence of conformal field theory, there is a one-to-one correspondence between the energy eigenvalues $E_n$ and the scaling dimensions of operators $\Delta_n$:
\begin{align} 
E_n = \frac{\hbar c \Delta_n}{R}
\end{align}
where $c$ is the velocity of `light', and the ground state has zero energy. The lowest energy levels correspond to operators originally studied by Wilson and Fisher \cite{WilsonFisher}, and they have $\Delta_n$ and their spacings of order unity. So for $E \sim \hbar c /R$, we have the many-body density of states $D(E) \sim R/(\hbar c)$. This small value of $D(E)$ at low $E$ is quite different from the SYK model and charged black holes. 

Let us now consider larger energies with $E \gg \hbar c/ R$. At sufficiently large energies, we expect chaotic behavior with eigenstate thermalization, and so the system should be characterized locally by a flat space theory at a temperature $T$. From the entropy density in (\ref{e2}) for $d=2$,  we obtain the total entropy
\begin{align} 
S(T) = (4 \pi R^2) \, a_2 k_B (k_B T /(\hbar c))^2  \,. \label{sta2}
\end{align}
We can now use familiar thermodynamic identities ($S=-\partial F/\partial T$, $F=E-TS$) to obtain the energy
\begin{align} 
E(T) = \frac{2 T}{3} S(T) \,,\label{sta3}
\end{align}
and then convert to the
microcanonical ensemble
\begin{align} 
S(E) = k_B (9 \pi a_2)^{1/3} \left( \frac{E R}{\hbar c} \right)^{2/3} \,.
\end{align}
Finally, from Boltzmann's relation we obtain the many-body density of states
\begin{align}
D(E) \sim \exp \left[  (9 \pi a_2)^{1/3} \left( \frac{E R}{\hbar c} \right)^{2/3} \right] \,.\label{dewf}
\end{align}
This shows that the spacing between energy levels for $E \gg \hbar c/ R$ is exponentially small in $E^{2/3}$, and (\ref{dewf}) should be compared with the exponentials (obtained from the large argument limit of the sinh) of $E^{1/2}$ in (\ref{de}) for the SYK model, and in (\ref{DEF}) for charged black holes. 
These large dimension operators of the Wilson-Fisher theory correspond to neutral black hole states, although their gravity theory is surely more complicated than Einstein gravity. It would be interesting to study such black hole states numerically by the methods of Ref.~\onlinecite{YinChen22}.

We can now see that a significant advantage of the SYK model was that its black holes states extended all the way down to the ground state, and this was an important reason for its simplicity. In contrast, the Wilson-Fisher theory is black hole-like only above some threshold energy, and the low-lying states are not exponentially dense.

When carried out for a conformal field theory in $d$ spatial dimensions, the above argument yields $D(E) \sim \exp \left[ b E^{d/(d+1)} \right]$, for some constant $b$, for black hole states \cite{Shaghoulian:2015kta,Belin:2016yll}. This does not match the SYK model $D(E) \sim \exp \left[ b E^{1/2} \right]$ at $d=0$ because of the important role of the Schwarzian-induced breaking of conformal symmetry: the state-operator correspondence does not apply to the SYK model.

\section{Discussion}

The main results for the common structure of the many-body densities of states of non-supersymmetric charged black holes and the SYK model have been obtained in Sections~\ref{sec:SYK} and \ref{sec:bh}, and were summarized in Fig.~\ref{fig:bhdos}. This connection shows that it is possible to realize Hawking's black hole entropy in a manner consistent with the standard quantum mechanics of many-body systems without supersymmetry. The match is at the level of the density of states coarse-grained over a few level spacings, and extends also to multi-point correlators of the density of states \cite{Saad:2019lba}. 

The precise energy levels of the SYK model in Fig.~\ref{fig:grisha}, of course, depend upon the particular realization chosen for the couplings $U_{ij;k\ell}$. 
We do not expect any such instance of the SYK model to described the ultimate microstructure of a realistic black hole. But numerous universal features of realistic black holes are indeed captured by the SYK model \cite{Bousso:2022ntt}, after coarse-graining over an exponentially small level spacing.

The reader may find the analogy with single-particle quantum chaos instructive. A particle moving in a quantum billiard defines a problem with no randomness in the Hamiltonian, analogous to a black hole. Nevertheless, the energy levels are chaotic, and their statistical properties are equivalent to those of a random matrix ensemble \cite{Bohigas,Altland_chaos}. For the black hole, the SYK model defines a random matrix in the many-body Fock space of size $2^N \times 2^N$ using of order $N^4$ random numbers: so the SYK matrix is quite sparse and has significant structure, and is very far from a fully random matrix with $2^{2N}$ independent elements.

Finally, we note that the SYK model was originally introduced \cite{SY93} 
as a toy model of the strange metal state of the cuprates. This connection has also undergone rapid development in recent years, and has led to realistic and universal model of strange metal behavior \cite{Patel:2022gdh,CLi24}, as reviewed in a companion article \cite{SSORE}.

\subsection*{Acknowledgements}

This article is based on lectures presented at the College de France, Paris in May-June 2022. I am grateful to Antoine Georges for his warm hospitality, and for many discussions and collaborations over the years. I also thank Matthew Heydeman, Luca Iliesiu, Upamanyu Moitra, Sameer Murthy, Leo Radzihovsky, Sandeep Trivedi, G.~Joaquin Turiaci, and Spenta Wadia for valuable discussions. This research was supported by the U.S. National Science Foundation grant No. DMR-2002850, and by the Simons Collaboration on Ultra-Quantum Matter which is a grant from the Simons Foundation (651440, S.S.).

\bibliography{mef.bib}      


\end{document}